\documentclass[a4paper,twoside,notoc,12pt]{JHEP3}
\usepackage{epsfig,multicol}

\newcommand{\nn}{\nonumber}
\newcommand{\beq} {\begin{equation}}
\newcommand{\eeq} {\end{equation}}
\newcommand{\beqa} {\begin{eqnarray}}
\newcommand{\eeqa} {\end{eqnarray}}
\newcommand{\mrm}[1] {{\mathrm{#1}}}
\newcommand{\mbf}[1] {{\mathbf{#1}}}

\newcommand{\ie}{{\it i.e.}}
\newcommand{\eg}{{\it e.g.}}
\newcommand{\cf}{{\it cf.}}
\newcommand{\etal}{{\it et al.}}

\newcommand\gev{~{\rm GeV}}
\newcommand{\as}{\alpha_s}
\newcommand{\lqcd}{\Lambda_{\rm QCD}}

\newcommand{\order}[1]{${\cal O}\left(#1 \right)$}
\newcommand{\morder}[1]{{\cal O}\left(#1 \right)}
\newcommand{\eq}[1]{(\ref{#1})}

\newcommand{\half}{{\scriptstyle \frac{1}{2}}}

\newcommand{\lsim}{\lesssim}
\newcommand{\gsim}{\gtrsim}

\newcommand{\zb}{\bar z}

\newcommand{\tr}[1]{ {\bf #1}_\perp}

\def\pit{P}         
\def\qt{k}          

\title{Subprocess Size in Hard Exclusive Scattering}
\author{Paul Hoyer$^1$, Jonathan T.\ Lenaghan$^{2,3}$, Kimmo
Tuominen$^{4,5}$ and Carsten Vogt$^4$\\
$^1$Department of Physical Sciences and
Helsinki Institute of Physics\\ \vspace{3mm}
POB 64, FIN-00014 Helsinki University, Finland\\
$^2$Department of Physics, University of Virginia\\\vspace{3mm}
382 McCormick Rd.,Charlottesville, VA 22903, USA\\
$^3$Niels Bohr Institute and $^4$Nordita\\ Blegdamsvej 17, DK-2100
Copenhagen, Denmark}

\preprint{May, 2004\\ HIP-2002-47/TH \\ NORDITA-2002-67 HE\\
\hepph{0210124}}

\abstract{The interaction region of hard exclusive hadron
scattering can have a large transverse size due to endpoint
contributions, where one parton carries most of the hadron
momentum. The endpoint region is enhanced and can dominate in
processes involving multiple scattering and quark helicity flip.
The endpoint Fock states have perturbatively short lifetimes and
scatter softly in the target. We give plausible arguments that
endpoint contributions can explain the apparent absence of color
transparency in fixed angle exclusive scattering and the
dimensional scaling of transverse $\rho$ photoproduction at high
momentum transfer, which requires quark helicity flip. We also
present a quantitative estimate of Sudakov effects.}


\begin{document}

\section{The dynamics of endpoint contributions}

In the Brodsky-Lepage (BL)
framework of exclusive scattering
\cite{bl:1980,Brodsky:1989pv}, the amplitude for a process $A+B \to C+D$
factorizes at large $t=(p_A-p_C)^2$ into a hard subprocess $a+b
\to c+d$ times distribution amplitudes $\phi_{a/A},\ldots$ for
each external hadron. Here $a$ represents the valence Fock state
of hadron $A$ (\eg, $a = uud$ for the proton). The distribution
amplitude is the valence Fock state amplitude at equal Light-Cone
(LC) time, integrated over the relative transverse momenta of the
partons up to a hard scale of $\morder{\sqrt{-t}}$. Hence all
hadrons involved in the scattering are in transversally compact
configurations. In the subprocess amplitude the momenta of the
partons in each hadron are effectively parallel, their relative
transverse momenta being negligible compared to the hard scale.

There are factorization theorems of various degree of rigour for QCD reactions. 
However, it should be kept in mind that for 
semi-exclusive processes and for hadronic projectiles in general, factorization is a 
strong assumption: no rigorous QCD proof exists. In this paper we qualitatively study 
which aspects of the data conform to general expectations when factorization is 
assumed. Our investigation builds on a physical picture rather than a precise 
formalism, and serves to indicate potential problems and possible solutions. This will 
hopefully guide the development of a more systematic formalism in future work.

The assumed factorization framework may indeed fail due to endpoint contributions in 
the integration over the longitudinal momentum
fractions of the quarks \cite{Brodsky:1989pv,ilr}.
The longitudinal momentum of a quark with fraction
$z \lsim \lqcd/p^+$ of its parent hadron momentum~$p^+$
(in a frame where the hadron moves fast along the $z$-axis)
is no larger than its transverse momentum.
Hence the quark is isotropically
distributed in momentum space and outside the light-cone
formed by the fast quarks of the subprocess. The virtualities of
subprocess propagators decrease with $z$, allowing hadron Fock
states of large transverse size to contribute near the endpoints.

The lifetime $\tau$ of a Fock state in a parent hadron of high
momentum $p^+= E+p^z$ is inversely proportional to the difference
between its energy and that of the hadron,
\beq \label{lifetime}
\frac{1}{\tau} \simeq \sum_i E_i - E \simeq \sum_i
\frac{k_{\perp i}^2 + m_i^2}{z_i \, p^+} - \frac{M^2}{p^+} \,,
\eeq
where $M$ is the mass of the parent. The second approximation between
the ordinary and light-cone energy differences is valid when
$z_i \, p^+=k_i^+ \gg k_{\perp i},m_i$ for all constituents~$i$ and
is thus {\em not} valid for Fock states in the endpoint region.
The lifetime of the endpoint Fock states is $\sim 1/\lqcd$,
which is short compared to the typical lifetimes of
\order{p^+/\lqcd^2} of Fock states where all constituents have
comparable momentum fractions~$z_i$.

The nature of endpoint dynamics is illustrated by Deep Inelastic
Scattering ($e \, p \to e \, X$, DIS). In the aligned jet (parton
model) regime the virtual photon with $q^+ \simeq 2 \, \nu$ splits
asymmetrically into a $q\bar q$ pair, such that $z_{\bar q} \sim
\lqcd^2/Q^2$ and $k_{\perp\bar q} \sim \lqcd$~\cite{bks,bhm}.
Thus the antiquark momentum $k_{\bar q}^+ = z_{\bar q} \, q^+
\simeq \lqcd^2/(m_N x_B)$ stays finite in the Bjorken limit. The
probability $\propto 1/Q^2$ of the asymmetric photon splitting
determines the scaling of the DIS cross section,
$\sigma_{tot}(\gamma^*p) \propto 1/Q^2$. The non-perturbative
scattering cross section $\sigma[(q\bar q)N] \sim 1/\lqcd^2$ of
the $q\bar q$ Fock state corresponds to the quark distribution
$f_{q/N}(x_B)$ in the $q^- \simeq 2 \, \nu$ (or Breit) frame.

The endpoint dynamics was also studied for quarkonium
hadroproduction, $\pi N \to J/\psi + X$ \cite{bhmt}. QCD
factorization breaks down when the quarkonium carries large
fractional momentum, $x_F \simeq 1- \lqcd^2/M_{J/\psi}^2$. In
this regime there is no hard scattering on a target parton (\eg,
in a subprocess such as $gg \to c\bar c$). Rather, a compact
Fock state in the pion  projectile fluctuates into an endpoint
state where nearly all momentum is carried by the heavy quark
pair. The light valence quarks have transverse momenta of
\order{\lqcd} and their soft, non-perturbative scattering in the
target liberates the heavy quarks, which then appear in the final
state.

According to Eq. \eq{lifetime} all endpoint configurations have
short lifetimes in spite of their large transverse size. Like
compact states they thus have a low number of constituents -- in
particular, comoving fields that have long formation times are
absent. Hence there is no enhanced forward radiation in scattering
processes.

The large size of endpoint configurations favors multiple
scattering in the target. This enhances their importance in
diffractive processes which require color singlet exchange. The
aligned jet configuration contributes at leading twist to
diffractive DIS, whereas multiple scattering of the compact,
symmetric $q\bar q$ configurations is power suppressed. Endpoint
configurations are also enhanced in scattering on nuclear
targets due to the increased importance of multiple scattering.
The nuclear dependence in effect measures the size of the
contributing Fock states. The apparent absence of color
transparency in large angle $e \, p \to e \, p$ \cite{ctep} and
$p \, p \to p \, p$ \cite{ctpp} scattering, with the target proton
embedded in a nucleus, may signal dominant endpoint contributions
in these processes\footnote{See Ref.~\cite{Sargsian:2002wc} for a
discussion and alternative explanations.}.

Quark helicity flip in hard photon and gluon interactions is
suppressed by a factor $m_q/k_\perp$. Helicity is therefore
conserved at leading twist in BL factorization. On
the other hand, the low $k_\perp \sim \lqcd$ of endpoint
constituents implies that quark helicity flip is not suppressed.
The relative importance of amplitudes with quark helicity flip
is thus another measure of endpoint contributions.

The LC energy difference in \eq{lifetime} diverges when any
fractional momentum $z_i \to 0$. This is the reason why
distribution amplitudes, which are defined at equal LC time,
vanish at the endpoints \cite{bl:1980,Brodsky:1989pv}. As we
emphasized above, however, target scattering is soft in the
endpoint regime, implying a breakdown of LC dominance and of
factorization into hard subprocess and distribution amplitudes.
For this dynamics it is more natural to use the difference of
ordinary energies in~\eq{lifetime} which stays finite (albeit large)
in the $z_i \to 0$ limit. This increases the importance of endpoint
contributions in convolution integrals.

In this paper we study two processes where data indicates that
endpoint contributions dominate. The perturbative QCD (PQCD)
estimate for $d\sigma/dt(\gamma \, p \to \pi^+ n)$, obtained from
the semi-exclusive process $\gamma \, p \to \pi^+ Y$
(Fig.~1)~\cite{BDHP:1998} using Bloom-Gilman duality~\cite{BG, CEBAF},
is two orders of magnitude below the data \cite{hoyer:2001}. In
section~2 we show that the transverse size of the  $\gamma \, u \to
\pi^+ d$ subprocess is effectively large, and that the color
transparency assumed in the semi-exclusive process is thus
likely to be violated.

Recent high energy data on $\rho^0$ and $\phi$ meson
photoproduction show a dominance of quark helicity flip out to
large momentum transfer $|t| \lsim 12
\gev^2$~\cite{Chekanov:2002}. In section~3 we study the properties
of the $\gamma \, g \to \rho\, g$ subprocess amplitude. The
amplitude for longitudinally polarized $\rho$ mesons (which
conserves quark helicity) vanishes for real external photons. The
amplitude for transversely polarized $\rho$'s (which dominates in
the data) has strongly enhanced endpoint contributions. Due to the
suppression of quark helicity flip in hard scattering only the
soft endpoint contributions can potentially explain the observed
dimensional scaling of the cross section. We discuss qualitatively
how the endpoint region might give rise to dimensional scaling.

\section{The size of $\gamma \, u \to \pi^+ d$}

\FIGURE[t]{\epsfig{file=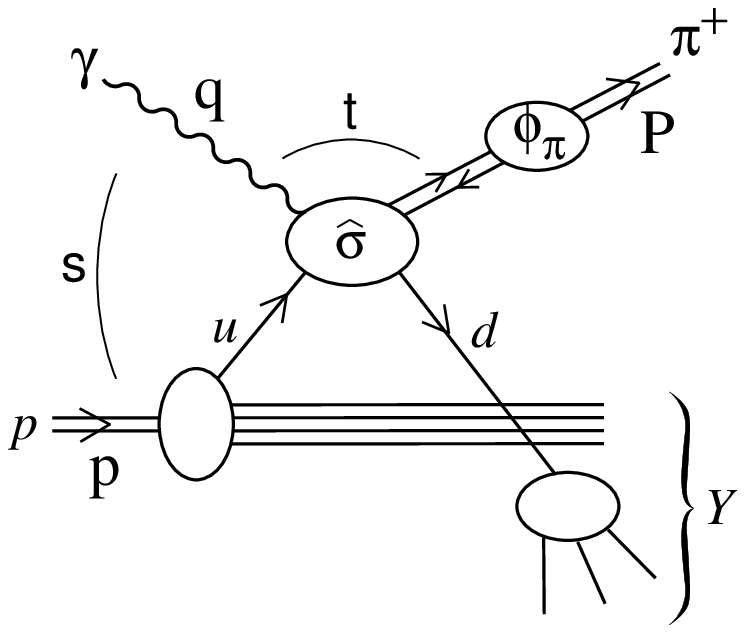,width=.6\textwidth}
\caption{Semi-exclusive scattering. In the limit (2.1) the cross
section factorizes into a hard subprocess cross section
$\hat\sigma$ times a target parton distribution.}}

Our study of the $\gamma \, u \to \pi^+ d$ process is motivated by
its role as a subprocess of semi-exclusive $\gamma \, p \to \pi^+
Y$ scattering. Although no rigorous proof exists, the applicability of factorization 
is often assumed. Then, in the kinematic limit
\beq \label{sexlimit}
s \gg -t, M_Y^2 \gg \lqcd^2 \,,
\eeq
where (\cf\ Fig.~1) $s=(q+p)^2 = E_{\rm CM}^2$ and the invariant
momentum transfer $t=(q-P)^2$, the semi-exclusive cross section
reads~\cite{BDHP:1998}
\beq \label{sexsigma}
\frac{d\,\sigma}{dt}(\gamma \, p \to \pi^+ Y) = \sum_{q=u,\bar d}
f_{q/p}(x)\, \frac{d\hat{\sigma}}{dt}(\gamma \, q \to \pi^+ q') \,.
\eeq
The fractional momentum of the struck quark is $x=-t/(M_Y^2-t)$
and the subprocess cross section is given by
($\hat s = xs$)
\begin{eqnarray}\label{subsigma}
      \frac{d\hat{\sigma}}{dt}(\gamma \, u \to \pi^+ d) =
          \frac{256\,\pi^2\,\alpha\,\as^2}{27\,\hat{s}^2\,|t|} \,
          (e_u - e_d)^2 \,
          \left[ \int_0^1 dz \, \frac{\Phi_\pi(z)}{z} \right]^2 \,.
\end{eqnarray}
Here $\Phi_\pi(z)$ is the pion distribution amplitude, \ie, its
Fock state wave function for a $u\bar d$ pair at short
transverse distance $\sim \morder{1/\sqrt{-t}}$, with the
$u$-quark carrying a fraction $z$ of the pion momentum. A color
singlet $u\bar d$ pair of small transverse size does not
rescatter in the proton target, giving the simple expression
\eq{sexsigma} for the semi-exclusive cross section.

There is as yet no data on $\gamma \, p \to \pi^+ Y$ in the kinematic
region~\eq{sexlimit}. Assuming that semi-exclusive processes obey
Bloom-Gilman duality, one may relate the $\gamma \, p \to \pi^+ Y$
cross section to the one for $\gamma \, p \to \pi^+ n$. However, the
measured $\gamma \, p \to \pi^+ n$ cross section is so large that
Bloom-Gilman duality would have to fail by two orders of magnitude
for the prediction of the semi-exclusive cross section to be
correct~\cite{hoyer:2001}.

In light of the recent experimental evidence for
Bloom-Gilman duality in inclusive reactions~\cite{CEBAF},
such a gross failure seems rather unlikely. A more plausible
explanation is that~\eq{sexsigma} is an underestimate of the true
$\gamma \, p \to \pi^+ Y$ cross section due to a lack of color
transparency.

We shall use the photon virtuality as a probe of the transverse
size of the subprocess \eq{subsigma}. The cross section is
independent of $Q^2$ when the size of the scattering region is
small compared to $1/Q^2$. We employ the asymptotic distribution
amplitude
\beq \label{asdist}
\Phi_\pi(z) = \frac{\sqrt{6}}{2}f_\pi\, z(1-z)
\eeq
($f_\pi\simeq 130$ MeV) and neglect quark masses ($m_q=0$). In
Fig.~2 we show the differential cross section
$d\sigma/dt(\gamma_\mrm{\, T}^*(Q^2) \, u \to \pi^+ d)$ (solid line)
for a transversely polarized virtual photon as a function the
dimensionless ratio $Q^2/|t|$. While the real photon cross
section given by~\eq{subsigma} is finite, its {\em slope} at
$Q^2=0$ is (as we shall see, logarithmically) infinite. Thus,
however big the momentum transfer $|t|$ is, the transverse size
of the photon scattering region remains large. For comparison we
also show (dashed line in Fig.~2) that the Compton scattering
$\gamma^* e \to \gamma \, e$ cross section is independent of $Q^2$
in the limit $s \gg |t|$, as expected due to the pointlike nature
of the photon.

\FIGURE[h]{\epsfig{file=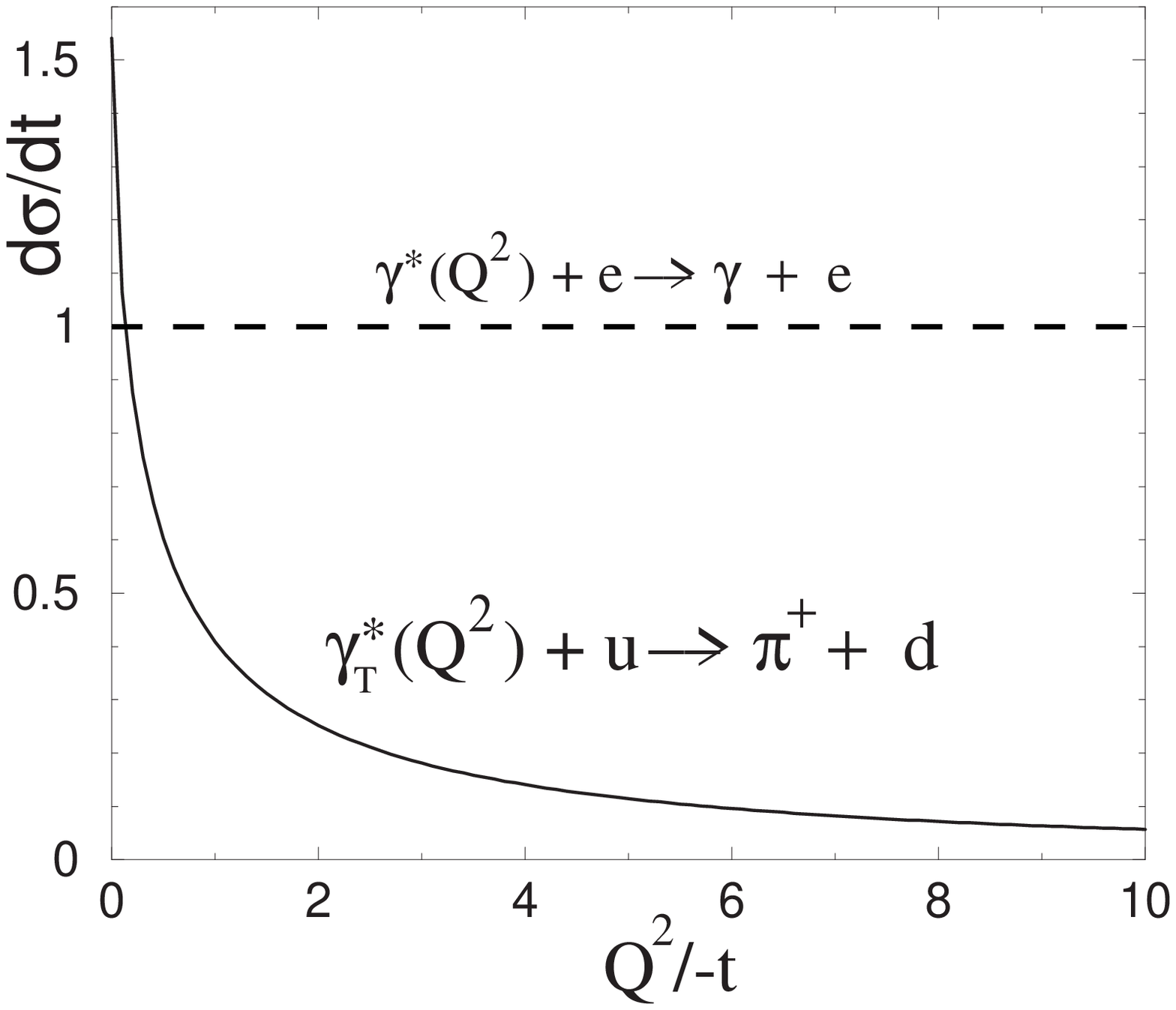,width=0.6\textwidth}
\caption{$d\sigma/dt(\gamma_\mrm{\, T}^*(Q^2)+u \to \pi^+ +d)$
for a transversely polarized photon as a function of $Q^2/|t|$
(solid line). For comparison, we show the corresponding plot for
Compton scattering, $\gamma^*(Q^2)+e \to \gamma +e$ (dashed line).
The normalization is arbitrary.}}

The divergent slope of the meson photoproduction cross section
has already been noted by the authors of~\cite{ginz:1996}. They
concluded that the onset of the perturbative regime for the
production of light vector mesons is at rather large momentum
transfers, well beyond $30 \gev^2$.

The strong $Q^2$ dependence of the $\gamma^* u \to \pi^+ d$
cross section is due to endpoint contributions, even though the
$z$-integral in \eq{subsigma} is not enhanced near $z=0,1$. This
may be seen from the expression for the scattering amplitude,
which is a convolution of the pion distribution amplitude
\eq{asdist} with the $\gamma^* u \to (u\bar d) + d$ subamplitude
${\cal H}_{\mu, \lambda\lambda'}^{\nu\nu'}$ (\cf\ Fig.~3a),
\beq \label{convamp}
{\cal M}_{\mu,\lambda\lambda'}^{AB}(\gamma^* + u \to \pi^+ + d) =
      \delta_{AB} \int_0^1 dz \,
       \frac{1}{\sqrt{2}} \, \Big[ {\cal H}_{\mu,\lambda\lambda'}^{+-}(z)
       - {\cal H}_{\mu,\lambda\lambda'}^{-+}(z) \Big] \,
\Phi_{\pi}(z) \,.
\eeq
Here $A,B$ are the color indices of the incoming $u$- and outgoing
$d$-quark, $\mu$ is the photon helicity, $\lambda\ (\lambda')$
is the incoming $u$-quark (outgoing $d$-quark) helicity and
$\nu\ (\nu')$ is the $u\ (\bar d)$ quark helicity in the pair
forming the $\pi^+$. For $m_u = m_d =0$ quark helicity is
conserved\footnote{Expressions for the helicity amplitudes for
$Q^2,m_q \neq 0$ are given in the Appendix.} and we shall only
consider the $\mu=+1, \lambda=\lambda'=-\half$ amplitude. In the
semi-exclusive limit~\eq{sexlimit}
($s \gg -t,Q^2$),
\beq \label{subamp}
      {\cal H}^{-+}_{+,--} =
          - \frac{2 \sqrt{2} \, e \, (4 \pi \as) \, C_F}{\sqrt{-t}} \,
           \Bigg[ \frac{e_u}{ z - \bar{z} \, Q^2/t }
          - \frac{ e_d \, \zb }{z \, (\zb - z \, Q^2/t) } \Bigg] \,,
\eeq
where $C_F=(N_c^2-1)/2N_c$ is
the color factor and $\zb=1-z$.
We make the following observations:
\begin{itemize}
\item[(a)] At $Q^2=0$ the amplitude is $\propto (e_u-e_d)/z$.
This endpoint behavior therefore arises both from the photon coupling
to the `slow' $u$-quark and to the `fast' $d$-quark. On the
other hand, the amplitude is finite for $z \to 1$ since the
$u$-quark helicity $\nu = -\half$ is opposite to that of the photon
helicity $\mu = +1$.
Thus the helicity flip between the projectile and
fast outgoing particles is minimized.

\item[(b)] $d{\cal H}/dQ^2 \propto e_u/z^2$ for $z \to 0$ at
$Q^2=0$. The $1/z^2$ behavior gives a logarithmic singularity in
the convolution \eq{convamp} when $\Phi_{\pi}(z) \propto z$.
This is the origin of the infinite slope in Fig.~2. We also note
that the singular contribution arises from the photon coupling
to the {\em slow} quark.

This is distinct from the well known Feynman endpoint mechanism, where
the photon couples to the {\em fast} quark.
\end{itemize}

\FIGURE[h]{\epsfig{file=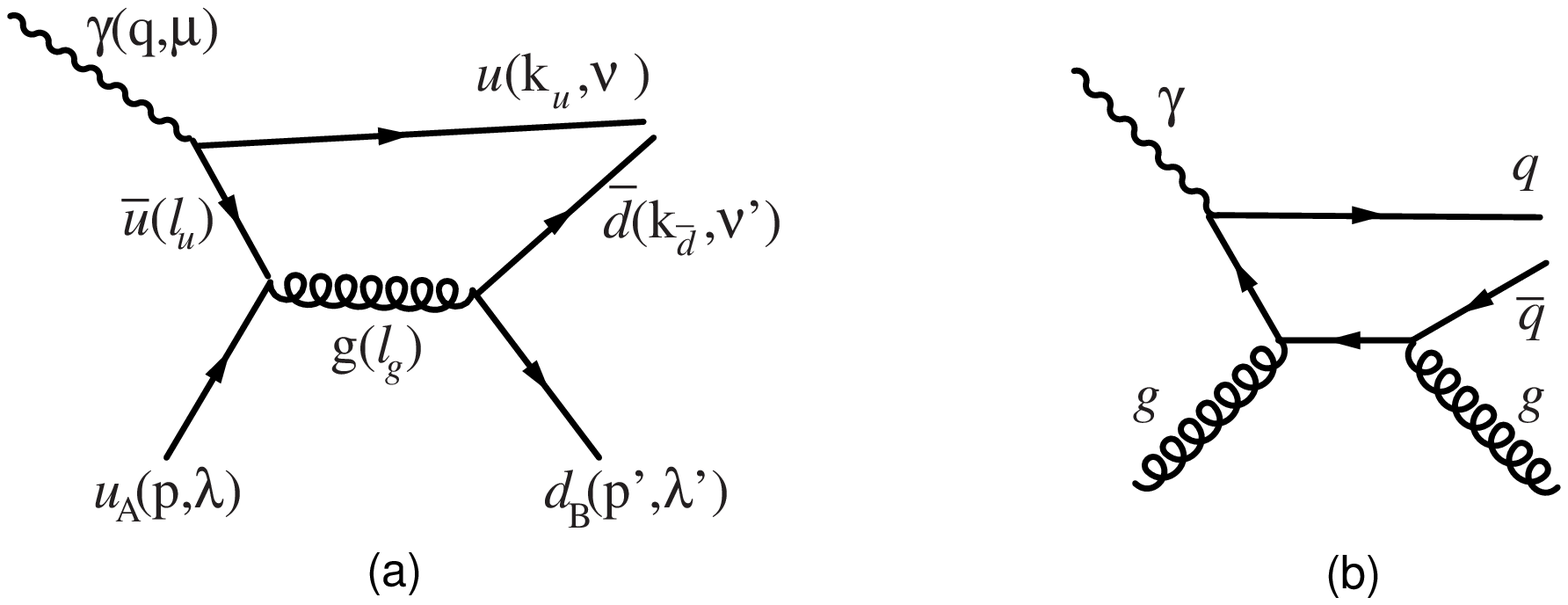,width=.8\textwidth}
\caption{(a) A diagram contributing to $\gamma+u \to u\bar d +
d$. (b) A diagram contributing to $\gamma+g \to q\bar q + g$.}}

The $Q^2$-sensitivity is due to the large transverse size of the
$\gamma \, u \to \pi^+ d$ process, as can be seen explicitly from
the kinematics (\cf\ Fig.~3a). We use the notation
$v=[ \, v^+,v^-,\mbf{v}_\perp \, ]$ where $v^\pm = v^0 \pm v^3$
for light-cone coordinates and take the photon momentum in the
negative $z$-direction so that\footnote{For notational convenience
we denote the (target) $u$-quark momentum by $p$ and the subprocess
energy by $s$ rather than $\hat s$.}
\beqa \label{momdef}
p&=&\sqrt{s} \, [ \, 1 \, , \, 0 \, , \, \mbf{0}_\perp \, ] \,,\nn\\
q&=&\sqrt{s} \, [ \, 0 \, , \, 1 \, , \, \mbf{0}_\perp \, ] \,,\nn\\
\\
k_u&=&
\left[\frac{(z \, \mbf{\pit}_\perp+\mbf{\qt}_\perp)^2}{z
\sqrt{s}} \, , \,
z \sqrt{s} \, , \, z \, \mbf{\pit}_\perp +
\mbf{\qt}_\perp\right] \,, \nn\\
k_{\bar d}&=& \left[\frac{(\bar
z \, \mbf{\pit}_\perp-\mbf{\qt}_\perp)^2}{\bar z \sqrt{s}} \, , \, \bar
z \sqrt{s} \, , \, \bar z \, \mbf{\pit}_\perp -
\mbf{\qt}_\perp\right] \nn
\eeqa
for on-shell massless quarks.
$\mbf{\pit}_\perp$ denotes the pion's transverse momentum relative
to the collision axis and $\mbf{\qt}_\perp$ is the relative
transverse momentum of its $u\bar d$ constituents.
The pion momentum $P = k_u+k_{\bar d}$
satisfies\footnote{If the $u$ and $\bar d$ quarks are not
on-shell (as would be the case in the pion) we have
$P^2=(\qt_\perp^2+z k_u^2+\zb k_{\bar d}^2)/z\zb$. Hence an
off-shellness of \order{\qt_\perp^2} is sufficient to keep the
pion on-shell even for $z\to 0,1$.} $P^2= \qt_\perp^2/z\bar z$,
where $\qt_\perp = $ \order{\lqcd},
and the momentum transfer $t=(q-P)^2 = -\pit_\perp^2$.

The virtualities of the internal quark and gluon lines in Fig.~3a are
\beqa \label{virtualities}
\ell_u^2 &=& (q-k_u)^2 = -z \pit_\perp^2- 2 \,
\mbf{\pit}_\perp\cdot\mbf{\qt}_\perp-\qt_\perp^2/z \,,\\
\ell_g^2 &=& (p+\ell_u)^2 = \bar zs +\ell_u^2 \,.
\eeqa
In the semi-exclusive limit \eq{sexlimit} we see that $\ell_u^2$
becomes sensitive to $\mbf{\qt}_\perp$ for
\beq \label{softz}
z \lsim \frac{\qt_\perp}{\pit_\perp} \sim \frac{\lqcd}{\sqrt{-t}} \,.
\eeq
Hence the subprocess is not transversally compact: the distance
between the photon absorption and gluon emission vertices in Fig.~3a
is given by the inverse of $\ell_{u\perp} =|z \,
\mbf{\pit}_\perp+\mbf{\qt}_\perp| = \morder{\lqcd}$ in the region
\eq{softz}. Moreover, BL factorization fails in this
endpoint region since the hard subamplitude depends on the relative
momentum of the quarks in the pion. These consequences of the
kinematics imply the $Q^2$-sensitivity of the process.

We can also see why the large transverse size, \ie,
the $Q^2$-sensitivity, arises only from the photon coupling to the slow
quark. When the $u$-quark is fast, \ie, for $z \simeq 1-
\lqcd/\sqrt{-t}$, the transverse distance $1/\ell_{u\perp} =
\morder{1/\sqrt{|t|}}$ while $\ell_g^2 \propto s\lqcd/\sqrt{-t}$
remains large since $s \gg -t$. Hence the distances between all
interaction vertices in Fig.~3a are short for $z \to 1$.

The fact that the $Q^2$ derivative of the subprocess amplitude
\eq{subamp} is more endpoint sensitive than the the amplitude
itself shows that a size measurement introduces inverse factors
of $z$ and $1-z$. Our result does not change the fact that the
$z$-integral of the leading twist cross section is flat (for the
asymptotic distribution amplitude \eq{asdist}), and thus is
dominated by compact configurations. However, rescattering in
the target will introduce dipole factors proportional to the
transverse size and cause the convolution integral to be
endpoint dominated. A failure of color transparency is thus a
likely reason for the large discrepancy with data found in
Ref.~\cite{hoyer:2001} for the $\gamma \, p \to \pi^+ n$ cross
section.

An analogous sensitivity to small photon virtualities can be
observed for the photon-pion transition form factor~\cite{DKV1},
$\gamma^*(Q^2) \, \gamma^{(*)}(Q'^2) \to \pi^0\,$: its rate of
change is logarithmically divergent as the ratio $Q'^2/Q^2 \to 0$.
At $Q'^2 = 0$ the pion transition form factor is given by the
same integral over the distribution amplitude as appears in the
cross section~\eq{subsigma}. The fact that the
$\pi^0$ is produced in isolation and color transparency thus is
not an issue may explain the phenomenological success
(see~\cite{Gronberg:1997fj} and references therein)
of the PQCD prediction in this case.

We have checked that the Sudakov effect does not change our
conclusions, by applying the modified factorization approach of
Ref.~\cite{bls} to the subprocess $\gamma \, u \to \pi^+ d$ in the
above kinematical limit. In our numerical calculation we employed
the asymptotic distribution amplitude~(\ref{asdist}), and two
parametrizations of the transverse momentum dependence of the
pion's LC wave function: a Gaussian of the form~\cite{jak:1993}
$\exp[- a_\pi^2\,k_\perp^2/(z\bar z)]$, with $a_\pi \simeq 0.86
\gev^{-1}$ being the transverse size parameter, and a Gaussian of
the form~\cite{BHL} $\exp[- \beta^2\,(k_\perp^2 + m^2_{\rm
eff})/(z\bar z)]$ with an effective mass $m_{\rm eff}=0.33\gev$
and $\beta \simeq 0.94 \gev^{-1}$. In both cases we found moderate
Sudakov corrections of about 5 -- 10\%, which shows that Sudakov
effects do not play a signi\-ficant role in the present
discussion. Similar qualitative and quantitative conclusions about
the impact of Sudakov corrections have been reached in
Refs.~\cite{jain:1995,DKV1}.

\section{Dimensional scaling with quark helicity flip in $\gamma
+p \to \rho+Y$}

\subsection{The experimental evidence}

The ZEUS collaboration recently published~\cite{Chekanov:2002}
data on $\rho$ photoproduction, $\gamma +p \to \rho+Y$, in the
semi-exclusive kinematics specified by Eq.~\eq{sexlimit}. The
data cover $80 < \sqrt{s} < 120$ GeV, $1.1 < |t| < 12\ \gev^2$
and is integrated over $x=-t/(M_Y^2-t) \gsim 0.01$.
The scattering is believed to be dominantly diffractive and,
due to the high value of $|t|$, to provide a testing ground for
the BFKL exchange mechanism~\cite{bfkl}. The hard
subprocess is then $\gamma + g \to \rho+g$, with the t-channel
containing a two-gluon ladder and the $\rho$ emerging via its
distribution amplitude according to BL factorization.

However, the data \cite{Chekanov:2002} pose a serious challenge
to this picture. Dimensional scaling predicts
\beq \label{rhoscaling}
\frac{d\sigma(\gamma \, g \to \rho \, g)}{dt}
\propto \frac{I_\rho^2}{|t|^n}
\eeq
with  $n=3$. Here $I_\rho$ is an integral over the $\rho$
distribution amplitude with dimension GeV. In contrast to the
quark exchange cross section \eq{subsigma} there is no factor
$s$ in the denominator of this gluon exchange cross section. The
data agree with dimensional scaling, giving $n=3.21\pm 0.04 \pm
0.15$ (in $\phi$ production the corresponding power is measured
to be $n=2.7 \pm 0.1 \pm 0.2$). Together with the fact that the
$\phi/\rho$ cross section ratio is consistent with the ratio 2/9
of the charge factors for $|t| \gsim 4\ \gev^2$, this suggests
that the $\gamma + g \to \rho+g$ process is hard and perturbative.

The upper part of the subprocess is shown in Fig.~3b. The quark
pair produced at the photon vertex scatters off the two gluons
and forms the vector meson via its distribution amplitude. The
vector meson is expected to be longitudinally polarized since
its quark and antiquark constituents have opposite helicities
due to helicity conservation at the photon and gluon vertices.
However, the ZEUS data show that the $\rho$ meson inherits (to a
good approximation and in the full $t$-range) the transverse
polarization of the incoming photon. In the BL factorization
framework this implies a quark helicity flip, incurring an
$m_q^2/|t|$ suppression factor in the cross section
\eq{rhoscaling}, which is then expected to scale with a power
$n=4$.

Thus, we are faced with a dilemma. The data obeys simple
dimensional scaling ($n=3$), is consistent with the quark
production process being hard ($\phi/\rho$ flavor symmetry
indicates insensitivity to $m_q$) and the semi-exclusively
produced $\rho$ meson carries both the momentum and the helicity
of the projectile. But these attractive features are mutually
inconsistent within the standard factorization
framework~\cite{bl:1980} of exclusive processes.

The authors of~\cite{ivanov:2000} consider the possibility
that the production of transverse vector mesons is due to a
non-perturbative, chiral-odd wave function of the photon, which
is proportional to the quark condensate. This contribution is
subleading at asymptotically large momentum transfers. In such
an approach the dimensional scaling observed~\cite{Chekanov:2002}
for $|t| \lsim 12 \gev^2$ would be accidental.

\subsection{The subprocess amplitudes}

To resolve the dilemma let us consider the structure of the
factorized $\gamma + g \to \rho+g$ amplitude. It is a convolution
of the quark pair production amplitude
${\cal G}(\gamma + g \to q\bar q+g)$ and
the $\rho$ distribution amplitude $\Phi_\rho^{\mu'}$.
For example, for transversely polarized $(\mu'= +1)\ \rho$ mesons,
\beq \label{convampg}
{\cal M}_{\mu,\lambda\lambda'}^{ab}(\gamma + g \to \rho+g) =
       \delta_{ab}\int_0^1 dz \, {\cal
G}_{\mu,\lambda\lambda'}^{++}(z) \, \Phi_{\rho}^+(z) \,.
\eeq
Here $a,b$ are gluon color indices, the
$\mu,\lambda,\lambda'$ indices of ${\cal G}$ are the helicities
of the photon, incoming and outgoing gluon, respectively, and
the upper indices refer to the $q$ and $\bar q$ helicities.

For simplicity we consider only the lowest order contribution to
${\cal G}(\gamma + g \to q\bar q+g)$ (\cf\ Fig.~3b). Higher
order diagrams build the gluon ladder and are important for
describing the $s$-dependence, but should not affect the
helicity structure of the upper vertex, which is our present
concern.

The transverse photon ($\mu=+1$) amplitude with
$\lambda=\lambda'= +1$ and no quark helicity flip is
\beqa \label{g-trans-long}
{\cal G}^{+-}_{+,++}(\gamma + g \to q\bar q+g) &=&
    -\frac{\sqrt{2} \, e e_q \, (4\pi\as)}{\sqrt{N_c}\, \sqrt{-t}}
    \frac{Q^2}{t} \,\nn\\
    &\times& \frac{2 \, z - 1}{\bar{z} \, ( z - \bar{z} \,
Q^2/t - m_q^2/t)( \bar{z} - z \, Q^2/t - m_q^2/t ) } \,\,.
\eeqa
This leading twist amplitude {\em vanishes} in
photoproduction ($Q^2=0$). The quark helicity flip amplitude
contributing to transverse $\rho$ production in \eq{convampg} is
at\footnote{Expressions for $Q^2\neq 0$
are given in the Appendix.} $Q^2=0$
\beqa
\label{g-trans-trans}
    {\cal G}^{++}_{+,++}(\gamma + g \to q\bar q+g) &=&
    - \frac{\sqrt{2}\, e e_q \, (4\pi\as)}{\sqrt{N_c}\, \sqrt{-t}} \,
      \frac{\sqrt{m_q^2/(-t)}}{z \bar{z} \, ( z -  m_q^2/t )
      ( \bar{z} - m_q^2/t)} \nn\\ &&\\
   &=& - \frac{\sqrt{2}\, e e_q \, (4\pi\as)}{\sqrt{N_c}\, \sqrt{-t}}
      \, \frac{\sqrt{m_q^2/(-t)}}{(z \bar{z})^2}\left[1+
      \morder{\frac{m_q^2}{t}}\right] \,. \nn
\eeqa
The factor $(z\zb)^2$ in the denominator enhances the endpoint
regions $z=0,1$ in the convolution \eq{convampg} causing a
(logarithmic) singularity in the $z$-integral for distribution
amplitudes which vanish linearly at the endpoints. This implies
a breakdown of factorization in semi-exclusive $\rho$
photoproduction.

\subsection{Endpoint behavior of the distribution amplitude}

\FIGURE[t]{\label{fig:one-gluon}
\psfig{file=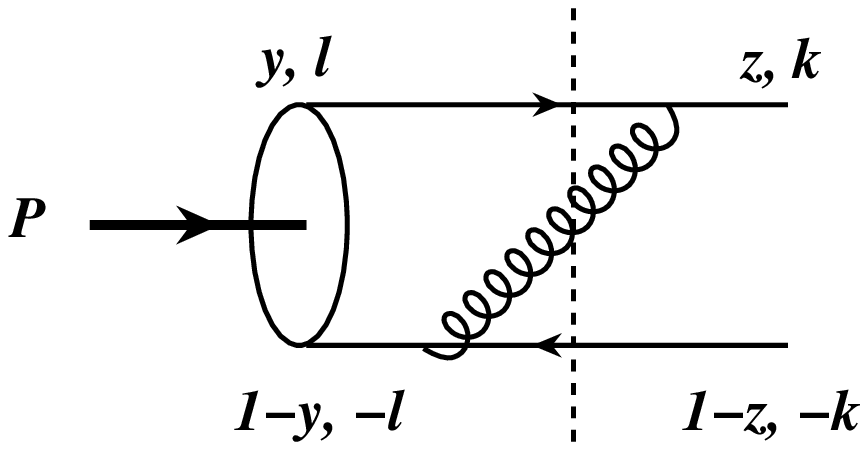,width=.4\textwidth}
\caption{A representative of the one-gluon exchange diagrams
contributing to the evolution equation in time-ordered LC
perturbation theory.}}

Even though factorization fails due to the strong enhancement of
the subprocess amplitude in the region where one of the produced
quarks carries nearly all the momentum, we may still see
qualitatively how the experimentally observed dimensional
scaling of the cross section can arise. As we emphasized above,
this must be due to the endpoint contributions, since for finite
values of $z$ the transverse $\rho$ cross section is suppressed
by $m_q^2/|t|$, giving $n=4$ in Eq. \eq{rhoscaling}.

The distribution amplitude $\Phi_{\rho}^{\mu'}(z)$ appearing in
the convolution \eq{convampg} is the valence Fock state amplitude
{\em at equal light-cone (LC) time}. Its endpoint behavior can be
determined from the evolution equation, which, at leading order,
corresponds to one-gluon exchange diagrams, \cf\
Fig.~\ref{fig:one-gluon}~\cite{bl:1980}. Thus, the wave function
can be written as
 \beq \label{eq:wavefct} \Phi_\rho(z,\mu_F) =
\int dy \int^{\mu_F} \frac{d^2 \tr{l}}{16\,\pi^3} \, \as(\mu_F) \,
\frac{\mathcal{S}(y,z,\tr{l},\tr{k})} {P^- - \sum_i p_i^-+ i
\epsilon} \, \Phi_\rho(y,\tr{l}) \,, \eeq where $P^-$ is the meson
LC energy and the sum is over all LC energies $p_i^-$ of the
intermediate $q \bar q g$ state. $\mathcal{S}$ is a momentum
dependent function which is finite at the endpoints. Hence, the
behavior for $z\to 1$ follows from the energy denominator of the
intermediate state, \beq P^- - \sum_i p_i^-\simeq -\frac{k_\perp^2
+ m_q^2} {(1-z) \, P^+} \quad \mathrm{for} \quad z \to 1 \,. \eeq
An analogous behavior can be found for $z \to 0$. Consequently,
$\Phi_\rho(z,\mu_F)$ vanishes at $z=0,1$ because the LC energy of
a parton with momentum fraction $z$ tends to infinity as $z \to
0$. However, as we saw in the previous section, the subprocess is
soft and therefore not light-cone dominated for $z \lsim z_s
\simeq \lqcd/\sqrt{-t}$. The LC energy is not relevant in this
region -- while the ordinary energy difference \eq{lifetime}
obviously remains finite and is of \order{\lqcd} as the
longitudinal momentum of a parton vanishes.

This suggests that $\Phi_{\rho}^{\mu'}(z)$ (effectively) does not
vanish at the endpoints. The $z$-integral is then linearly divergent
at $z=0,1$. Truncating the integration region
where the subprocess becomes soft we get
\beq
\int_{z_s}^{1-z_s} \frac{\Phi_{\rho}^{\mu'}(z)}{z^2\zb^2}
\propto \frac{\sqrt{-t}}{\lqcd}\, \Phi_{\rho}^{\mu'}(0) \,.
\eeq
Thus we {\em gain} a factor $\sqrt{-t}$ in the amplitude, which
restores dimensional scaling, implying $n=3$ in Eq.~\eq{rhoscaling}.
As already remarked above, here we only give qualitative arguments
for the observed scaling. In order to predict the normalization of
the cross section a more detailed analysis is necessary, which is
however beyond the scope of the present work.

\section{Summary}

In this paper we have attempted to give a physical picture able to explain some 
qualitative aspects of the data on meson production at large momentum
transfer $|t|$ by real, transversely polarized photons. Also, we have pointed out why 
some expectations based on the common assumption of factorization in exclusive PQCD 
studies seem to fail. Namely, we saw that the scattering is
likely to be endpoint dominated and thus involve $q\bar q$ pairs
of large transverse size. Hence it appears that highly
asymmetric Fock states, where one quark carries nearly all the
momentum, do have a significant overlap with hadron wave functions.
This would explain the apparent absence of color transparency
in $e \, A \to e \, p \, (A-1)$ \cite{ctep} and $p \, A \to p \,
p \, (A-1)$~\cite{ctpp}.

Endpoint contributions are enhanced in photoproduction since the
wave function of transverse photons does not vanish at $z=0,1$.
This also prevents factorization of the transverse photon
amplitude in deeply virtual meson production at high $Q^2$ and
low $|t|$ \cite{cfs}. The longitudinal photon wave function
$\Psi_\gamma^{\mu=0}(z) \propto z(1-z)$ favors contributions
from $q\bar q$ pairs of small transverse size $\sim 1/Q$. The
color transparency observed in $\gamma^*(Q^2)+N \to \rho
+N$~\cite{Adams:1994bw,Arneodo:1994id,Ackerstaff:1998wt} agrees
with this.

Due to the short lifetime of the endpoint states their dynamics
has many of the attributes of hard scattering, despite their large
transverse size and soft scattering in the target. We have found
that the Sudakov form factor numerically leads to a suppression of
no more than about 10\%, and is thus of minor importance.
Furthermore, we have presented qualitative arguments suggesting
that endpoint contributions may explain the dimensional scaling of
the ZEUS data. The endpoint states are not color transparent, nor
do they preserve quark helicity. These features make it possible
to identify their contribution to exclusive processes.

\acknowledgments

It is a pleasure to thank Stan Brodsky and Jim Crittenden for many
helpful discussions and comments on the manuscript. This work was
begun while PH was employed by Nordita, and has been supported by
the European Commission under contract HPRN-CT-2000-00130 and by
the Academy of Finland under project 102046.


\appendix

\section{Appendix}

\subsection{Quark production amplitudes}

We present the helicity amplitudes $\cal{H}^{\nu\nu'}_{\mu,\lambda
\lambda'}$ for the subprocess $\gamma^*(\mu) \, u(\lambda)
\to u(\nu) \bar{d}(\nu') +
d(\lambda')$ for
$Q^2, \, m_q \neq 0$. For transversely polarized photons we find
\begin{eqnarray}
\label{qsubamp_t-l}
   {\cal H}^{+-}_{+,++} = - {\cal H}^{-+}_{-,--} &=&
   \frac{2 \, \sqrt{2} \, e \, (4 \pi \as) \, C_F}{\sqrt{-t}} \,
   \nn \\ &\times& \Bigg[ \frac{e_u \,z}{\bar{z} \,
   ( z - \bar{z}\,Q^2/t - m_q^2/t ) }
   - \frac{ e_d }{ ( \bar{z} - z\,Q^2/t - m_q^2/t ) } \Bigg] \,.
\end{eqnarray}
The amplitudes ${\cal H}^{+-}_{-,++}$ and ${\cal H}^{-+}_{+,--}$ are
obtained from ${\cal H}^{+-}_{+,++}$ and ${\cal H}^{-+}_{-,--}$,
respectively, by exchanging $z \to \bar{z}, \, e_u \leftrightarrow e_d$
and reversing the overall sign. For longitudinally polarized photons
we get
\begin{eqnarray}
\label{qsubamp_l-l}
   {\cal H}^{+-}_{0,++} = {\cal H}^{-+}_{0,--} =
   \frac{4 \, e \, (4 \pi \as) \, C_F}{\sqrt{-t}} \, \left[
   \frac{ e_u \, \sqrt{Q^2/(-t)} }{ z - \bar{z} \, Q^2/t - m_q^2/t}
   +\frac{ e_d \, \sqrt{Q^2/(-t)} }{ \bar{z} - z \, Q^2/t - m_q^2/t}
    \right] \,. \quad
\end{eqnarray}
The quark helicity flip amplitudes read
\begin{eqnarray}
\label{qsubamp_t-t}
   {\cal H}^{++}_{+,--} = {\cal H}^{--}_{-,++} =
   -\frac{2 \, \sqrt{2} \, e \, (4 \pi \as) \, C_F}{\sqrt{-t}} \,
   \left[\frac{e_u \, \sqrt{m_q^2/(-t)}}{z \, \bar{z} \,
   ( z - \bar{z}\,Q^2/t - m_q^2/t ) } \right] \,.
\end{eqnarray}
For ${\cal H}^{++}_{+,++}$ and ${\cal H}^{--}_{-,--}$ we have
to make the replacements $z \to \bar{z}, \, e_u \to e_d$ in
the above respective amplitudes. All other helicity amplitudes
vanish in the kinematical limit~\eq{sexlimit}. The helicity
non-flip amplitudes agree with those given in Ref.~\cite{huang:2000}
in the limit of large c.m. energies and when appropriate replacements
for the charge factors are made.\footnote{In Ref.~\cite{huang:2000}
the production of flavor-neutral mesons is considered. Also note
that the normalization of the subprocess amplitudes is different
than in our case.}
We note that in the helicity flip amplitudes~(\ref{qsubamp_t-t})
the photon couples only to the quark whose helicity is flipped.
According to \eq{virtualities} the gluon virtuality in Fig. 3a
is of order $s$. Hence to leading order in the limit \eq{sexlimit}
the quark helicity can flip only at the photon vertex.

\subsection{Gluon production amplitudes}

The helicity amplitudes for the subprocess
$\gamma^*(\mu) + g(\lambda) \to q(\nu) \, \bar{q}(\nu') + g(\lambda')$
are denoted by $\cal{G}^{\nu\nu'}_{\mu,\lambda\lambda'}$.
For transversely polarized photons we find
\begin{eqnarray}
\label{gsubamp_-t-l}
   {\cal G}^{+-}_{+,++} = - {\cal G}^{-+}_{-,++} &=&
   -\frac{\sqrt{2} \, e e_q \, (4 \pi \as) }
   {\sqrt{N_{\rm c}} \, \sqrt{-t}} \, \frac{Q^2}{t} \nn \\
   &\times& \frac{2\,z - 1}{ \bar{z} \, ( z - \bar{z} \, Q^2/t -
   m_q^2/t )\,  ( \bar{z} - z \, Q^2/t - m_q^2/t ) } \,,
\end{eqnarray}
and for longitudinally polarized photons
\begin{eqnarray}
\label{gsubamp_l-l}
   && {\cal G}^{+-}_{0,++} = {\cal G}^{-+}_{0,++} =
     \frac{2\, e e_q \, (4 \pi \as) }{\sqrt{N_{\rm c}} \, \sqrt{-t}} \,
     \frac{\sqrt{Q^2/(-t)}\, (1 - Q^2/t)}
     { (z - \bar{z} \, Q^2/t - m_q^2/t)\,
     ( \bar{z} - z \, Q^2 - m_q^2/t)} \,.
\end{eqnarray}
The amplitudes
${\cal G}^{-+}_{+,++}$ and ${\cal G}^{+-}_{-,++}$ are obtained
from ${\cal G}^{+-}_{+,++}$ and ${\cal G}^{-+}_{-,++}$
respectively by replacing $z \to 1-z$.
For the quark helicity flip amplitudes we obtain
\begin{eqnarray}
\label{gsubamp_t-t}
   {\cal G}^{++}_{+,++} = {\cal G}^{--}_{-,++} &=&
   -\frac{\sqrt{2} \, e e_q \, (4 \pi \as)}{\sqrt{N_{\rm c}} \,
\sqrt{-t}}
   \,\frac{\sqrt{m_q^2/(-t)}}{z\, \bar{z}} \nn \\
   &\times& \frac{1 - Q^2/t}{ ( z - \bar{z} \, Q^2/t - m_q^2/t ) \,
   ( \bar{z} - z \, Q^2/t - m_q^2/t)} \,.
\end{eqnarray}
For the above combination of quark and photon helicities, the set of
amplitudes with negative gluon helicities is identical, \ie,
\begin{equation}
   {\cal G}^{\nu\nu'}_{\mu,--}={\cal G}^{\nu\nu'}_{\mu,++} \,.
\end{equation}
All other helicity amplitudes vanish in the limit~\eq{sexlimit}.
Again, we find agreement with the results given in
Ref.~\cite{huang:2000}.


\end{document}